\documentstyle[11pt,moriond,epsfig]{article}

\def\lsim{\raise0.3ex\hbox{$<$\kern-0.75em\raise-1.1ex\hbox{$\sim$}}}
\def\gsim{\raise0.3ex\hbox{$>$\kern-0.75em\raise-1.1ex\hbox{$\sim$}}}

\def\noi{\noindent}

\def\be{\begin{equation}}
\def\ee{\end{equation}}
\def\bea{\begin{eqnarray}}
\def\eea{\end{eqnarray}}

\begin{document}
\vspace*{4cm}
\title{DIS DATA AND THE PROBLEM OF SATURATION
IN SMALL-$x$ PHYSICS.}

\author{A. CAPELLA$^a$, E.G. FERREIRO$^b$, A.B. KAIDALOV$^c$ AND 
\underline{C.A. SALGADO.}$^a$}

\address{$^a$Laboratoire de Physique Th\'eorique, U. de Paris XI,
B\^atiment 210, F-91405 Orsay Cedex, France.\\
$^b$Dep. de F\1sica de Part\1culas, U. de Santiago de 
Compostela,E-15706 Santiago de Compostela, Spain.\\
$^c$ITEP, B. Cheremushkinskaya ulitsa 25,
117259 Moscow, Russia.}

\maketitle\abstracts{
New experimental data from HERA explore the region of very small $x$ and
small and moderate $Q^2$. We study the role of unitarity corrections in the
description of these data. For that, we propose a model separating the
small and large components of the $q\bar q$ fluctuation. This model gives a
unified description of total and diffractive production in $\gamma^*p$
interactions. In this model
unitarity corrections are controlled by diffraction data.
}

One of the main experimental facts discovered at HERA is the fast
growth of parton densities as energy increases, or equivalently as $x\to 0$. 
Taking $\sigma^{tot}\sim s^{\alpha(0)-1}$ ($F_2\sim x^{-\alpha(0)+1}$)
values of $\Delta\equiv \alpha(0)-1$ in the range $0.1\div 0.5$ have been 
reported. These values depend on the virtuality $Q^2$
\footnote{In Regge phenomenology $\alpha(0)$ is the intercept of the exchanged
object, a pomeron at high energy, this is why two different
pomerons, a hard and a soft one, have been proposed in order to 
explain this feature. Unitarity effects, however, make also $\Delta_{eff}$
to depend on $Q^2$ and $x$, as in our case \cite{ckmt}.}. 
A behavior like this would violate unitarity
given by the Froissart bound ($\sigma\lsim (\log s)^2$ as $s\to\infty$).
Unitarity is restored in Regge models by allowing for multiple
exchange of Regge trajectories. We present a model \cite{lowx} 
which takes into account
all multiple pomeron exchanges in eikonal approximation. These unitarity
corrections are controlled by diffraction.
In parton language, the high density of partons (mainly gluons)
at small values of $x$ makes the gluon fusion to become important,
stopping the increase of the density. 
In any case, unitarity corrections are given by non-linear terms, the importance
of which grows with energy. 
These non-linearities would eventually give a saturation of partonic
densities \cite{satur} 
at small enough values of $x$ (the actual value depending, in general, on the
value of $Q^2$) and different models
try to understand the onset of this effects \cite{lowx} \cite{models}.
In experiments
with nuclei, parton densities are a factor $\sim A^{1/3}$ larger,
so, saturation starts at larger values of $x$ than in experiments with 
nucleons \cite{nucleos}. 

\section{The model.}

The description of the $lp$ collision in the laboratory frame is very
appropriate to include unitarity effects. In this frame, the virtual photon
$\gamma^*$ emitted by the $l$ fluctuates into a 
$q\bar q$ pair. This system suffers then multiple 
interactions with the proton.
We propose a separation in two components depending
on the transverse size $r$ of the $q\bar q$. A small component S for
$r<r_0$,
where the pQCD result, $\sigma^{q\bar qp}\sim r^2$, is used,
and a large L component, for $r>r_0$, 
described by Regge phenomenology. $r_0$ has been taken as a free parameter
and its value turned out to be small, $r_0$=0.2 fm. 
Unitarity corrections
to both components are given by multiple scattering in a generalized eikonal 
approach which includes  triple pomeron interaction. The total $\gamma^*p$
cross section is $\sigma^{tot}=\sigma^{tot}_S+\sigma^{tot}_L$ with

\be
\sigma_S^{tot}=4\int d^2b\int_0^{r_0}d^2r\int dz |\psi(r,z)|^2
{1-\exp(-C\chi_S(x,Q^2,b,r))\over 2C},
\label{eq1}
\ee

\be
\sigma_L^{tot}=4g_L^2(Q^2)\int d^2b {1-\exp(-C\chi_L(x,Q^2,b))\over 2C},
\label{eq2}
\ee

\noi
where $|\psi(r,z)|^2$ and $g_L^2(Q^2)$ describe the $\gamma^*-q\bar q$
coupling (see ref. \cite{lowx}), 
$C$ is a parameter to take into account the dissociation of the
proton. The eikonals are

\be
\chi_S(r, b, s, Q^2) = {\chi_{S0}(r,b,s,Q^2) \over 1 + a \ \chi_3(b,s,Q^2)} ,
\label{eq3}
\ee

\be
\chi_L(s,b,Q^2) = {\chi_{L0}^P(b, \xi) \over 1 + a \ \chi_3(s,b,Q^2)} +
\chi_{L0}^f (b,
\xi) \quad .
\label{eq4}
\ee

\noi
Where
\be
\label{eq5}
\chi_{S(L)0}^k(b, \xi) = {C_{S(L)}^k \over \lambda_{0k}^{S(L)} (\xi)} \exp \left (
\Delta_k \xi - {b^2 \over 4 \lambda_{0k}^{S(L)}(\xi )} \right ) \quad ,
\ee

\noi and

\be
\label{eq6}
\Delta_k = \alpha_k(0) - 1 \quad , \quad \xi = \ell n {s + Q^2 \over s_0 + Q^2}
\quad ,
\quad \lambda_{0k}^{S(L)} = R_{0kS(L)}^2 + \alpha '_k \ \xi \quad .
\ee

With $a=0$, the model described above is a standard quasi-eikonal with 
Born terms given by pomeron plus $f$ exchanges ($k=P, f$). Note that the
contribution of the $f-$exchange to the S component is very small \cite{lowx} 
and has been neglected. In eq. (\ref{eq5}), the constants $C_L^P$ and
$C_L^f$ determine respectively the residues of the pomeron and
$f$-reggeon exchanges in the $q\bar q$-proton interaction. In 
contrast, due to the pQCD result $\sigma^{q\bar qp}\sim r^2$ mentioned
above, the corresponding coupling $C_S^P$ is taken to be proportional
to $r^2$ (see the discussion in section 2). 

In eqs. (\ref{eq3}) and (\ref{eq4}), $a=g_{pp}^P(0)r_{PPP}(0)/16\pi$,
where $g_{pp}^P(0)$ and $r_{PPP}(0)$ are the proton-pomeron coupling 
and the triple pomeron coupling respectively, both at $t=0$. The expresion
of $\chi_3$ can be found in \cite{lowx}. The denominator in eqs. (\ref{eq3})
and (\ref{eq4}) correspond to a resummation of triple pomeron branchings (the
so-called fan diagrams).
The values of the
parameters can be found in ref. \cite{lowx}. It is important that
we have fixed the pomeron intercept
$\alpha_P$=1.2, and are the unitarity corrections which 
make the effective intercept
to depend on $x$ and $Q^2$.

The diffraction cross section 
is given by non-linear terms $\chi_{S,L}^n$, $n>1$ and $\chi_{S,L}^i
\chi_3^j$, $i,j>0$
in the expansion of (\ref{eq1}) and (\ref{eq2}). 
We call the former S and L contributions to diffraction and the lastest
PPP contributions: $\sigma_{\gamma^*p}^{(diff)}=
\sigma_S^{diff}+\sigma_L^{diff}+\sigma_{PPP}$. The expressions of these
components can be found in \cite{lowx}. 


\section{Results and conclusions.}

\begin{figure}
\begin{center}
\psfig{figure=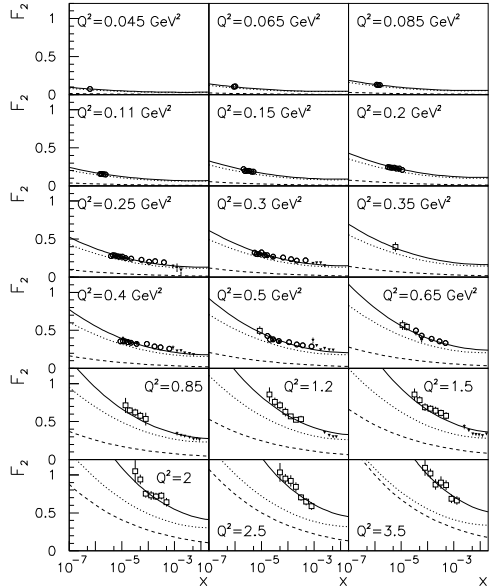,height=9cm}
\end{center}
\caption{
$F_2(x,Q^2)$ as a function of $x$ for different values of $Q^2$ compared
with experimental data from H1 1995 \protect\cite{h195} 
(open squares), ZEUS 1995
\protect\cite{zeus1995} (black circles), E665 \protect\cite{e665} 
(black triangles) 
and ZEUS BPT97 \protect\cite{zeus1997} (open circles). Dotted curve corresponds
to the $L$ contribution, dashed one to the
$S$ contribution and solid one to the total
$F_2(x,Q^2)$ given by the model.}
\label{fig1}
\end{figure}

\begin{figure}
\begin{center}
\psfig{figure=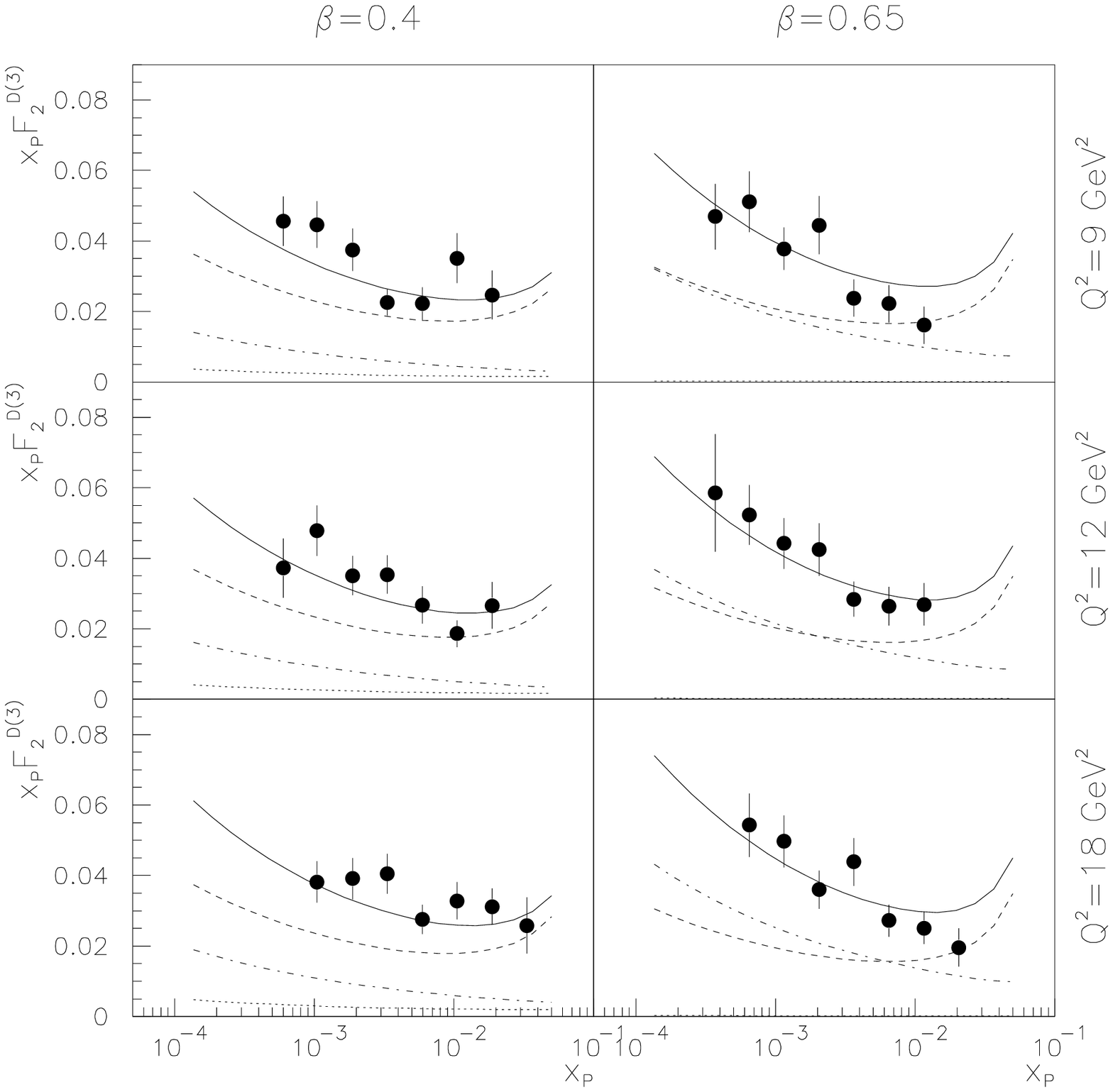,height=7cm}
\hspace*{2em}
\psfig{figure=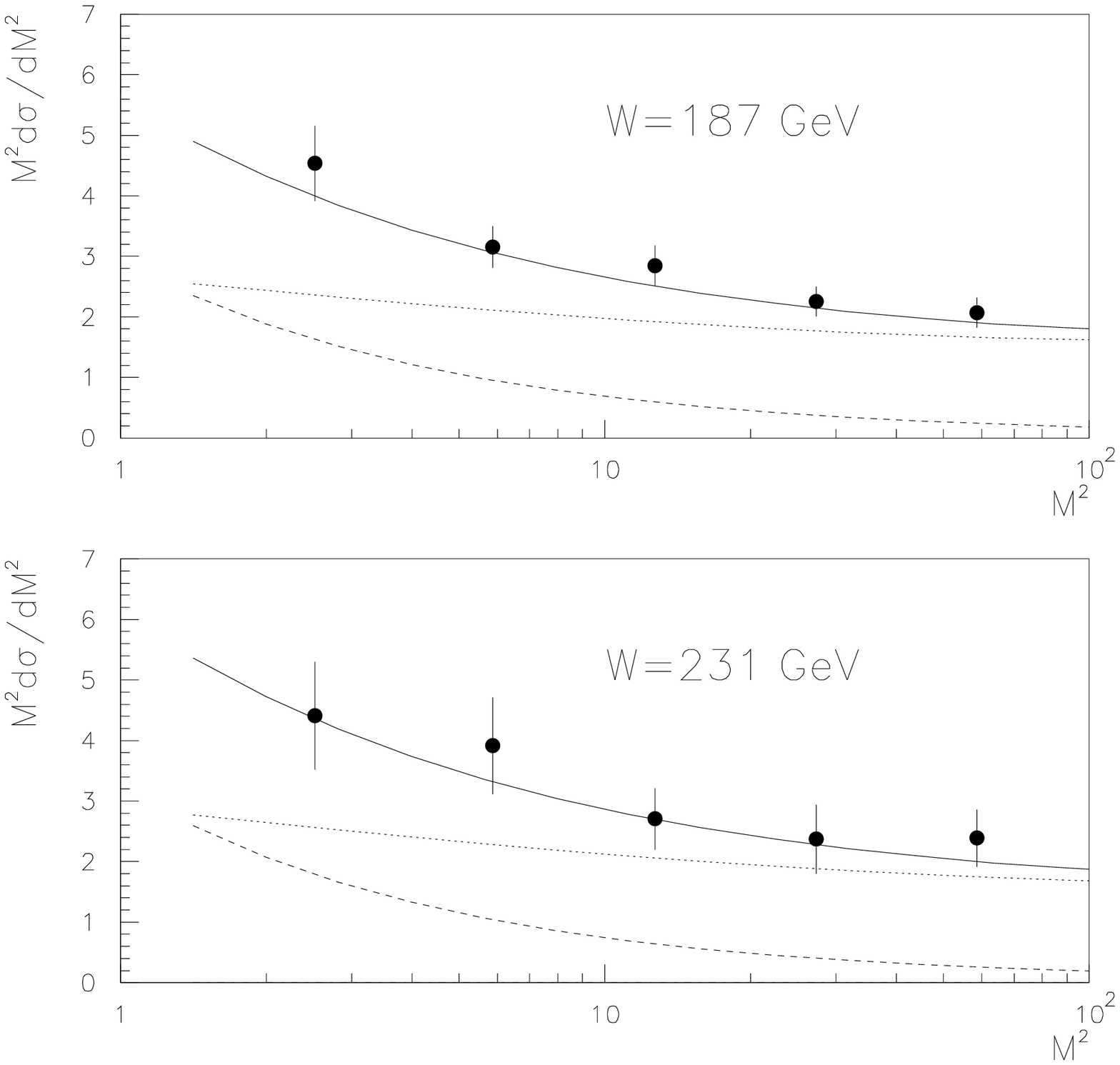,height=7cm}
\end{center}
\caption{Diffraction: 
$x_PF_{2D}^{(3)}$ as a function of $x_P$ for different $Q^2$
and
$\beta$ (left) data are from \protect\cite{h1diff} and 
diffractive photoproduction cross-section for $W$=187 and 231 GeV as
a function of $M^2$ from \protect\cite{photo} (right) 
compared with the model (solid lines).
The curves correspond to
PPP (dotted), L (dashed) and
S (dotted-dashed) contributions.
}
\label{fig2}
\end{figure}

In Fig. \ref{fig1} we compare our results with $F_2(x,Q^2)=Q^2/
(4\pi^2\alpha_{em})\sigma^{tot}(x,Q^2)$ 
data for different values of $Q^2$ from 0.045 GeV$^2$
to 3.5 GeV$^2$. It is clear from the figure that the S component
is negligible for small values of $Q^2$ 
and would eventually become bigger
than L one at large values of $Q^2$. In Fig. \ref{fig2} comparison with
diffraction is done for $x_PF_2^{D(3)}=Q^2/(4\pi^2\alpha_{em})\sigma^{diff}$
for different values of $Q^2$ and with diffractive 
photoproduction cross section.

In summary, we have developed a two components model 
that takes into account unitarity
corrections by multiple scattering of the $\gamma^*$
in a generalized eikonal approach which includes triple
pomeron. This unitarity corrections are controlled by the ratio $\sigma^{diff}/
\sigma^{tot}$, so, we have performed a joint fit of total and diffractive
data in lepton-proton collisions. The unitarity corrections for the
two components have different behaviors. At moderate values of $Q^2$,
$r^2\sim 1/Q^2$ and $\chi_S\sim 1/Q^2$, while $\chi_L$ is $Q^2$ independent. 
This makes unitarity corrections more important for the L than for the S 
component (in agreement with the fact that L part contribution to 
diffraction is
larger than S one). In fact, the mean number of collisions is only $\sim$ 1.1
for the S part at present energies, so the contribution to
unitarity corrections of the terms in powers of $1/Q^2$ (higher twist terms)
is less than 5\% at $Q^2\lsim$ 4 GeV$^2$. This results can
be extended to larger values of $Q^2$ by taking into account QCD evolution
with initial conditions given by the model \cite{qcdevol}. 
Finally, saturation will be reached when eq. (\ref{eq1}) and (\ref{eq2})
get the $(\log s)^2$ behavior. This will require energies much larger
than present ones \cite{saturnovo}.

\section*{Acknowledgments}
This work is partially supported by NATO grant PSTCLG-977275 and also
grants RFBR 00-15-96610, 0015-96786, 01-02-17095, 01-02-17383.
E.G.F. and C.A.S. thank Ministerio de Educaci\'on y
Cultura of Spain for financial support.
Laboratoire de Physique Th\'eorique is Unit\'e Mixte de Recherche - 
CNRS - UMR n$^{\circ}$ 8627.


\section*{References}
\begin{thebibliography}{99}

\bibitem{ckmt} A.~Capella, A.~Kaidalov, C.~Merino and J.~Tran Thanh Van,
Phys.\ Lett.\ B {\bf 337} (1994) 358; A.~B.~Kaidalov, C.~Merino and 
D.~Pertermann,
hep-ph/0004237.

\bibitem{lowx} A.~Capella, E.~G.~Ferreiro, A.~B.~Kaidalov and C.~A.~Salgado,
Phys.\ Rev.\ D {\bf 63} (2001) 054010; Nucl.\ Phys.\ B {\bf 593} (2001) 336.

\bibitem{satur} L. V. Gribov, E. M. Levin and M. G. Ryskin, Phys. Rep. 
{\bf 100} (1983) 1; 
E. Laenen and E. Levin, Ann. Rev. Nucl. Part. {\bf 44}, 199 (1994);
A. B. Kaidalov, Surveys High Energy Phys. {\bf 9}, 143 (1996);
A. H. Mueller, hep-ph/9911289.


\bibitem{models} 
E.~Iancu, A.~Leonidov and L.~McLerran, hep-ph/0102009;
A. H. Mueller, Nucl. Phys. {\bf B437} (1995) 107; 
K. Golec-Biernat and M. W\"usthoff, Phys. Rev. {\bf D59} (1999) 014017;
L L. Frankfurt, W. Kopf and M. Strikman, Phys. Rev. {\bf D57} (1998) 512;
A. L. Ayala, M. B. Gay Ducati and E. M. Levin, Phys. Lett. {\bf B388}
(1996) 188;
E. Gotsman, E. Levin and U. Maor, Nucl. Phys. {\bf B493}, 354 (1997).

\bibitem{nucleos}
L. Mc Lerran and R. Venugopalan, Phys. Rev. {\bf D49}
(1994) 2233, 3352; 
J. Jalilian-Marian et al., Phys. Rev. {\bf D59} (1999) 014014, 034007;
Yu. V. Kovchegov and A. H. Mueller, Nucl. Phys. {\bf B529} (1998) 451;
A. Capella, A. Kaidalov, J. Tran Thanh Van, Heavy Ion Phys. {\bf 9}
(1999) 169;  
N.~Armesto and C.~A.~Salgado, hep-ph/0011352. 

\bibitem{h195} C. Adloff {\it et al} (H1 Collaboration), Nucl. Phys. 
{\bf B497}
(1997) 3.

\bibitem{zeus1995} J. Breitweg {\it et al} (ZEUS Collaboration), Phys. Lett.
{\bf B407} (1997) 432.

\bibitem{e665} M. R. Adams {\it et al} (E665 Collaboration), Phys. Rev.
{\bf D54} (1996) 3006.

\bibitem{zeus1997} J. Breitweg {\it et al} (ZEUS Collaboration), 
Phys.\ Lett.\ B {\bf 487} (2000) 53.

\bibitem{h1diff} C. Adloff {\it et al} (H1 Collaboration), Z. Phys. {\bf C76},
 613 (1997).

\bibitem{photo}
C. Adloff {\it et al} (H1 Collaboration), Z.Phys. {\bf C74}, 221 (1997).

\bibitem{qcdevol} A.~Capella, A.~B.~Kaidalov and C.~A.~Salgado,
in preparation.

\bibitem{saturnovo} A.~Capella, A.~B.~Kaidalov and C.~A.~Salgado,
in preparation.

\end{thebibliography}
\end{document}